\documentclass[prl,twocolumn]{revtex4-1}

	\usepackage[english]{babel}
	\usepackage[utf8]{inputenc}

	\usepackage[T1]{fontenc}
	\usepackage[sc]{mathpazo}
	\usepackage{dsfont}
	\usepackage{mathrsfs}
	\usepackage[mathcal]{euscript}
	\usepackage{microtype}
	\usepackage{comment}
		
	\usepackage{color,graphics}
	\usepackage{amsmath}
	\usepackage{bm}
	\usepackage{amssymb}
	\usepackage{textcomp}
	\usepackage{indentfirst}
	\usepackage{enumerate}
	\usepackage{caption}
	\usepackage{subcaption}
	\usepackage{array}
	\usepackage{amsthm,dsfont}
	\usepackage{hyperref}
	\usepackage{cleveref}
	\usepackage{mathtools}

	\theoremstyle{definition}	
	\theoremstyle{definition}	
	\theoremstyle{plain}	
	\theoremstyle{definition}	
	\theoremstyle{plain}	

	\DeclareMathAlphabet{\mathpzc}{OT1}{pzc}{m}{it}

	\newcommand{\inteiros}{\ensuremath{\mathbb{Z}}}
	
	\newcommand{\unit}{\ensuremath{\mathds{1}}}

	\newcommand{\adj}[1]{\ensuremath{#1^*}}

	\newcommand{\order}{\ensuremath{\mathcal{O}}}

	\newcommand{\pint}[2]{\ensuremath{\langle #1,#2\rangle}}
	
	\newcommand{\der}[2]{\ensuremath{\frac{d}{d#2}#1}}

	\newcommand{\ket}[1]{\ensuremath{|#1\rangle}}
	\newcommand{\conj}[1]{\ensuremath{\overline{#1}}}

	\newcommand{\dd}{\,\ensuremath{\text{d}}}

	\renewcommand{\Re}{\operatorname{Re}}
	\renewcommand{\Im}{\operatorname{Im}}

\begin{document}

\title{Phase factors of periodically driven two-level systems}
\author{Marcela Muniz Gontijo}
\author{João C. A. Barata}
\affiliation{Department of Mathematical Physics, University of São Paulo,
São Paulo, Brazil}


\begin{abstract}

Using a perturbative solution for a periodically driven two-level quantum
system, we show how to obtain phase factors for both a two-level quantum
system and two two-level quantum systems non-interacting and interacting. 
The method is easily implemented by numerical routines and presents the
advantage of being stable for long-time periods. We furthermore explore
the possibility of implementing a quantum phase gate using the
perturbative solution.
\end{abstract}

\maketitle

\section{introduction}

The study of geometric phases has attracted significant interest since it
was shown that they could be used to process quantum
information \cite{ref:zanardi} and, due to its geometric properties,
they present an inherent resilience to fluctuation errors in the control
parameters. The experimental implementations of geometric phase in the
context of quantum computation, sometimes refered to as geometric quantum
computation (GQC), has been fruitful \cite{ref:jones,ref:anandanreview}. Nevertheless, to
obtain the expression for the geometric phase acquired by a two-level
quantum system, many works implement the rotating wave
approximation (RWA) \cite{ref:ekert,ref:sjoqvist,ref:jones}. As every
approximation, the RWA has its realm of validity and applicability that
has been extensively studied \cite{ref:blochsiegert, ref:stevenson,
ref:bonacci, ref:frasca, ref:spiegelberg}.

In this work, we consider the evolution of a two-level quantum system
driven by periodic fields. Instead of the RWA, we use the solution of
the Schrödinger equation obtained in \cite{ref:barata, ref:cortez}
(see also
\cite{ref:baratawre,ref:barataFormalQPsolutions,ref:baratacortez-ii,ref:gentilecortezbarata})
to compute the total, dynamical and geometric phase for a two-level
quantum system and two two-level quantum systems. Since the solution
used is uniformly convergent in time, the expressions for the phases
present a robustness when long-time periods are considered. We first
make a brief overview of the perturbative method developed in
\cite{ref:barata, ref:cortez}. The phases of a two-level quantum
system are then obtained using the perturbative expansion. The
discussion is extended to two two-level quantum systems,
non-interacting and interacting. In each case, we present the
expressions for calculating each phase factor. Finally, we obtain the
phase factors for the composite two two-level quantum system with a
delta interaction. We show that for a specific choice of parameters,
it is possible to build a phase shift gate.

\section{Description of the model and methods}

Let us start by considering a system with the following Hamiltonian:
\begin{equation}
	H_1(t) = \epsilon\sigma_3 - f(t)\sigma_1,
\label{eq:h1}
\end{equation}
where $\epsilon$ is a real constant and $f(t)$ is a periodic function
of time with frequency $\omega >0$. Let us consider a rotation of
$\pi/2$ around the $y$-axis, denoted by $R_y(\pi/2)$, and the
Schrödinger equation on this new rotated frame is given by
\begin{equation}
	i\der{\psi_2(t)}{t} = H_2(t)\psi_2(t),
\label{eq:schroh2}
\end{equation}
where 
\begin{equation}
\psi_2(t) = R_y(\pi/2)\psi(t) = \exp(-i\pi\sigma_2/4)\psi(t) \label{eq:psi2}
\end{equation}
and 
\begin{equation}
	H_2(t) = \epsilon\sigma_1 + f(t)\sigma_3.
\label{eq:h2}
\end{equation}
The Hamiltonian \eqref{eq:h1} can be interpreted as describing a system
with a Hamiltonian independent of time $\epsilon\sigma_3$ subjected to a
time-dependent perturbation $-f(t)\sigma_1$. The later is responsible for
transitions between the two states of the system.

The method developed in \cite{ref:barata} and \cite{ref:cortez} is
valid for small $\epsilon$ and periodic $f$. The quasi-periodic case
was analysed in \cite{ref:gentilecortezbarata}.  It consists in
writing a perturbative expansion in $\epsilon$ for the time evolution
operator. This method has proven to have the following advantages: the
series expansion are uniformly convergent in time, the expression
obtained for the time evolution operator is given in terms of series
and so are easily implementable in numerical calculations and they can
be employed for any periodic function. The uniform convergence is of
great importance, since it means the results lead to stable numerical
calculations and therefore allows the study of long-time behaviour
of the observable quantities of the system.

It was shown in \cite{ref:barata} that the time evolution operator $U(t)$
for the system described by \eqref{eq:h2} can be written as
\begin{equation}
	U(t) = \left(\begin{array}{cc} 
								R(t)(1+ig_0S(t)) & -i\epsilon R(t)S(t) \\
								-i\epsilon \conj{R(t)} \conj{S(t)} &
								\conj{R(t)}(1-i\conj{g_0}\conj{S(t)})
								\end{array}\right).
\label{eq:ubarata}
\end{equation}
where $R(t)$ and $S(t)$ are given by
\begin{equation}
	R(t) = e^{-i\Omega t} \sum_{m\in\inteiros} R_m e^{im\omega t}
\label{eq:rtbarata}
\end{equation}
and
\begin{equation}
	S(t) = \sigma_0 +  e^{2i\Omega t} \sum_{m\in\inteiros} S_m e^{im\omega
	t}.
\label{eq:stbarata}
\end{equation}
$R_m$ and $S_m$ are coefficients of the Fourier expansion of $R(t)$
and $S(t)$, respectively. Together with the Rabi frequency $\Omega$
and the constants $g_0$ and $\sigma_0$, they can all be obtained from
rather complex but convergent power series expansions in $\epsilon$,
involving the the Fourier coefficients of $f$ and its frequency
$\omega$.  See \cite{ref:cortez} as well as
\cite{ref:barata,ref:baratawre,ref:barataFormalQPsolutions,ref:baratacortez-ii}
for explicit formulas and examples. Sometimes we will refer to the
matrix elements of $U(t)$, for example, $U_{11}(t)=R(t)(1+ig_0S(t))$
and $U_{12}(t)=-i\epsilon R(t)S(t)$.

As done in \cite{ref:cortez}, we implemented numerically the method
developed there for a perturbation of the form
\begin{equation}
	f(t) = F_0 + A\cos(\omega t),
\label{eq:fbarata}
\end{equation}
where $F_0$ is a real number, $A$ and $\omega$ is the amplitude and the
frequency of the periodic perturbation, respectively.

Following the directions of the original paper, the method was applied to
several values of $\omega$ and $\epsilon$, the
former ranging from $1.0$ to $10.0$ and the later from $0.01$ to $0.40$.
For all these values, the unitarity test
was sufficiently satisfactory, since the error is bounded by $3\times
10^{-3}$ in one specific case (for $\omega=1.0$ and $\epsilon=0.40$), but
for most cases, is bounded by $10^{-5}$ or even $10^{-10}$.

\section{Total, dynamical and geometric phases}

We now show the calculations of the total, dynamical and geometric phases
for the two-level system considered. The total phase of the system is
simply given by
\begin{equation}
	\phi_{tot}(t) = \arg \pint{\psi(0)}{\psi(t)},
\label{eq:totphaset}
\end{equation}
and the dynamical phase $\alpha_{dyn}$ is given by
\begin{equation}
\alpha_{dyn}(t) = i\int_0^t \pint{\psi(t')}{\dot{\psi}(t')}\text{d}t' ,
	\label{eq:dynphaset}
\end{equation}
where $\psi(0)$ and $\psi(t)$ are the state vectors of the system at the
initial instant of time and for an instant of time $t$, respectively. The
dot indicates derivation relative to time. The geometric phase
$\gamma_{geo}$ is simply the difference between the total and dynamical
phases:
\begin{equation}
	\gamma_{geo}(t) = \phi_{tot}(t)-\alpha_{dyn}(t).
\label{eq:geophaset}
\end{equation}
We note that the phase factors are functions of time, since they are
defined by the evolution of the state vector $\psi(t)$. 

When performing the following calculations, we shall consider the state
vector correspondent to the rotated Hamiltonian given by \eqref{eq:psi2}.
The resulting expressions become
\begin{align}
	\phi_{tot}(t) &= \arg\{ \Re U_{11}(t) +i(-2\Re(\conj{\alpha}\beta)\Im
	U_{11}(t) \nonumber \\
	&\quad + 2\Im(\conj{\alpha}\beta)\Re U_{12}(t) + (2|\alpha|^2-1)\Im
	U_{12}(t))\} \label{eq:totrot}
\end{align}

and
\begin{align}
	&\alpha_{dyn}(t) = |\alpha|^2\left(-\Im \int_0^t a_{11}(t')dt' +
	i\Re\int_0^t a_{12}(t')dt'\right) \nonumber \\
	&\quad - 2i\Re(\conj{\alpha}\beta) \Re \int_0^t a_{11}(t')dt' -
	2i\Im(\conj{\alpha}\beta) \Im \int_0^t a_{12}(t')dt' \nonumber\\
	&\quad + |\beta|^2 \left(-\Im \int_0^t a_{11}(t')dt' -i\Re
	\int_0^t a_{12}(t')dt'\right),
\label{eq:dynphasea}
\end{align}
where $a_{11}(t)$ and $a_{12}(t)$ are matrix elements of the product of
$\adj{U}(t)$ and $\dot{U}(t)$:
\begin{equation}
	\adj{U}(t)\dot{U}(t) = \left(\begin{array}{cc}
																a_{11}(t) & a_{12}(t) \\
																-\conj{a}_{12}(t) & \conj{a}_{11}(t)
																\end{array}\right).
\label{eq:ajdudotu}
\end{equation}
The expression for the dynamical phase involves integrations over time
of the expansions. Although there are lots of integration routines,
using them in the highly oscillatory functions that constitute the
expansions often results in a large error due to the routine. Thus,
the integrations were carried out analytically term by term in the
Fourier expansions and then implemented numerically.

The previous expressions determine the total and dynamical phase for the
system for any instant of time. Next, it is necessary to define the
instant of time that is physically meaningful to the calculations of the
phase acquired by the system. One
could argue that the appropriate instant of time would be the ``natural''
frequency of the system, characterised by the Rabi frequency $\Omega$.
But we must recollect the nature of the geometric phase, that is, the
phase acquired over the course of the evolution of the system resulted
from the geometrical properties of the parameter space of the
Hamiltonian. In our case, the parameter space is two-dimensional, with
each dimension associated to the parameters $A$ and $\omega$ in
\eqref{eq:fbarata}. So, if we consider a cyclic evolution on the parameter
space and a fixed amplitude $A$ of the external field, the relevant
instant of time is precisely 
\begin{equation}
	t_\omega = \frac{2\pi}{\omega}.
\label{eq:tw}
\end{equation}
Therefore, the expressions \eqref{eq:totrot}, \eqref{eq:dynphasea} and
\eqref{eq:geophaset} for the respective total phase, dynamical phase and
geometric phase of the system are taken at $t_\omega$. Next, we present
some results of our calculations for the phase factors of the system as
graphical representations. Without loss of generality, we considered the
initial state vector to be $\psi(0)=\ket{0}$, that is, the state vector
is initially aligned with the $z$-axis. The calculations were performed
for values of $\epsilon$ ranging from $0.01$ to $0.40$ with steps of
$0.01$; and values of $\omega$ ranging from $1.0$ to $10.0$ with
steps of $0.5$.

\begin{figure}[!htp]
	\centering
	\includegraphics[width=.35\textwidth]{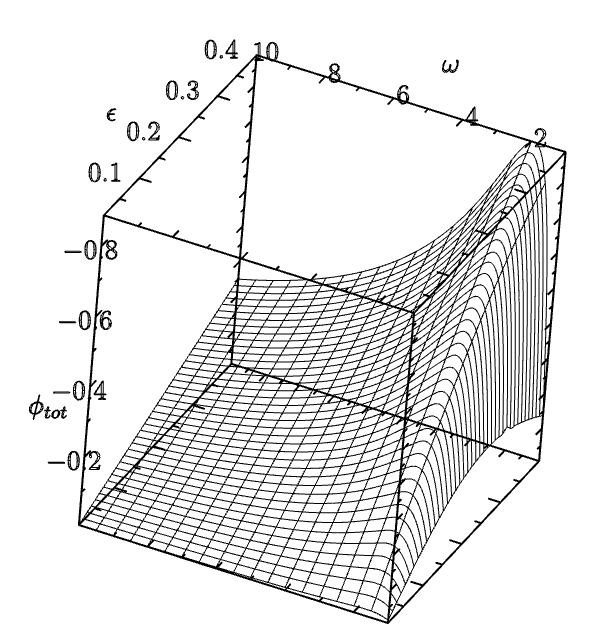}
	\caption{\small{Graphical representation of the total phase
	$\phi_{tot}$ as a function of $\omega$ and $\epsilon$.}}
\label{fig:totphase3d}
\end{figure}

As previously stated, the numerical implementation of the total phase was
easily accomplished. We note that since the total phase is defined as
an argument, there was no need to test if the numerical function had
relevant imaginary parts due to built-in machine errors. Figure
\ref{fig:totphasexe} shows the relation between the values of the total
phase and the parameter $\epsilon$ and Figure \ref{fig:totphase3d}
presents a three-dimensional representation of the total phase as a
function of $\omega$ and $\epsilon$. We can see that the absolute value of
the total phase is proportional to the value of $\epsilon$. According to
the interpretation of \eqref{eq:h1} in which $\epsilon$ is the energy gap
between the two eigenstates of $\sigma_3$, we can say that the total
phase is proportional to this gap. Moreover, we note that as the value of
$\omega$ increases, the rate in which the total phase increases with
$\epsilon$ decreases. In other words, the value of $\omega$ modulates the
curve $\phi_{tot} \times \epsilon$. Figure \ref{fig:totphasexw} shows
graphs of the total phase as a function of $\omega$ with fixed values of
$\epsilon$. The same behaviour observed in Figure \ref{fig:totphasexe} is
present in Figure \ref{fig:totphasexw}, but in this case, the value of
$\epsilon$ modulates the curve $\phi_{tot}\times\omega$ in the following
way: as $\epsilon$ increases, the curve gets more accentuated. It is also
notable that for $\omega$ around $2.0$, the absolute value of the total
phase is maximised. 
\begin{figure}[!htp]
	\centering
\begin{subfigure}[t]{.23\textwidth}
	\includegraphics[width=1.0\textwidth]{{{totphasexe}.pdf}}
	\caption{\small{Total phase $\phi_{tot}(t_\omega)$ as a function of
	$\epsilon$ for fixed $\omega=2.0$ (full line), $5.0$ (dashed line) and
	$10.0$ (dotted line).}}
	\label{fig:totphasexe}
\end{subfigure}
\hspace{.01\textwidth}
\begin{subfigure}[t]{.23\textwidth}
	\includegraphics[width=1.0\textwidth]{{{totphasexw}.pdf}}
	\caption{\small{Total phase $\phi_{tot}(t_\omega)$ as a function of
	$\omega$ for fixed $\epsilon=0.01$ (full line), $0.10$ (dashed line)
	and $0.40$ (dotted line).}}
	\label{fig:totphasexw}
\end{subfigure}
\caption{\small{Total phase plotted as a function of $\epsilon$ and
$\omega$.}}
\label{fig:totphase2d}
\end{figure}

\begin{figure}[!htp]
	\centering
	\includegraphics[width=.4\textwidth]{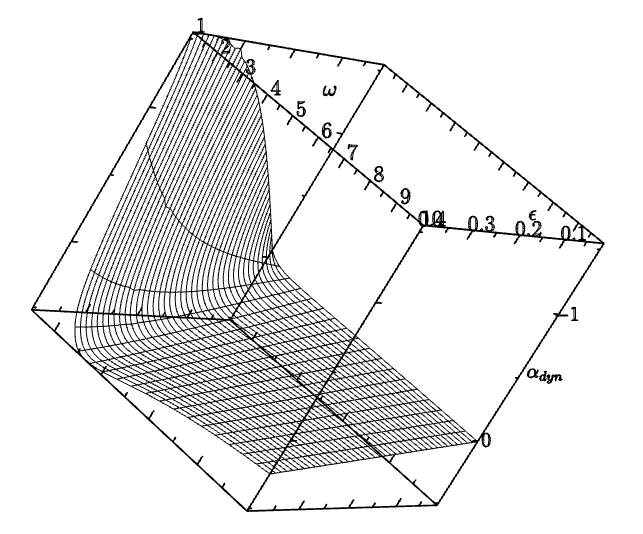}
	\caption{\small{Graphical representation of the dynamical phase
	$\alpha_{dyn}$ as a function of $\omega$ and $\epsilon$.}}
\label{fig:dynphase3d}
\end{figure}

The numerical implementation of the dynamical phase is not as
straightforward as that of the total phase, since it involves several
integrations over time (equation \eqref{eq:dynphasea}). These
integrations, as we said before, were done analytically and then
numerically implemented. The dynamical phase is expected to be real,
but the expansions in our implementation are truncated, so we tested
if the imaginary part of the dynamical phase had relevant
contributions. The imaginary parts equal zero within the machine
accuracy.  The relation between the dynamical phase and the values of
$\omega$ has a particular behaviour: for $\omega=1.0,1.5,2.0,2.5$ the
curve $\alpha_{dyn}\times\epsilon$ resembles a parabola and for higher
values the curve resembles a linear function. Figure
\ref{fig:dynphasexe} shows the dynamical phase as a function of
$\epsilon$ for some fixed values of $\omega$. Figure
\ref{fig:dynphasexw} shows the curve $\alpha_{dyn}\times\omega$ for
some values of $\epsilon$. We can see that, similar to Figure
\ref{fig:totphasexw}, $\epsilon$ seems to modulate the curve and there
is a value of $\omega$ that maximises $\alpha_{dyn}$, but this value
shifts according to the value of $\epsilon$. Figure
\ref{fig:dynphase3d} shows a three-dimensional representation of the
dynamical phase as a function of $\omega$ and $\epsilon$.
\begin{figure}[!htp]
	\centering
\begin{subfigure}[t]{.23\textwidth}
	\includegraphics[width=1.0\textwidth]{{{dynphasexe}.pdf}}
	\caption{\small{Dynamical phase $\alpha_{dyn}(t_\omega)$ as a function of
	$\epsilon$ for fixed $\omega=2.0$ (full line), $5.0$ (dashed line) and
	$10.0$ (dotted line).}}
	\label{fig:dynphasexe}
\end{subfigure}
\hspace{.01\textwidth}
\begin{subfigure}[t]{.23\textwidth}
	\includegraphics[width=1.0\textwidth]{{{dynphasexw}.pdf}}
	\caption{\small{Dynamical phase $\alpha_{dyn}(t_\omega)$ as a function of
	$\omega$ for fixed $\epsilon=0.01$ (full line), $0.10$ (dashed line)
	and $0.40$ (dotted line).}}
	\label{fig:dynphasexw}
\end{subfigure}
\caption{\small{Dynamical phase plotted as a function of $\epsilon$ and
$\omega$.}}
\label{fig:dynphase2d}
\end{figure}

\begin{figure}[!htp]
	\centering
	\includegraphics[width=.45\textwidth]{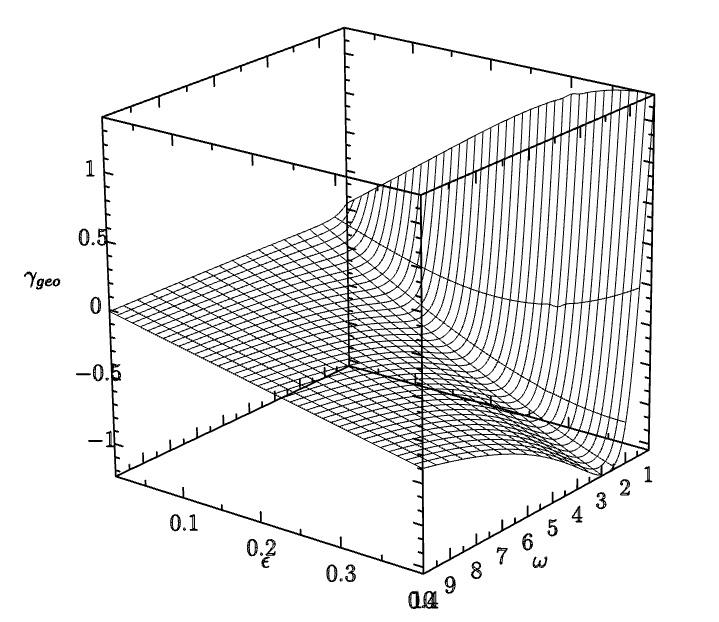}
	\caption{\small{Graphical representation of the geometric phase
	$\gamma_{geo}$ as a function of $\omega$ and $\epsilon$.}}
\label{fig:geophase3d}
\end{figure}

A similar behaviour of the total phase is observed for the geometric
phase in Figures \ref{fig:geophasexe} and \eqref{fig:geophasexw}:
the absolute value
of the geometric phase increases as $\epsilon$ increases, the curve
$\gamma_{geo}\times\omega$ is modulated by $\epsilon$ and it presents a
value of $\omega$ that maximises the absolute value of the geometric
phase. Figure \ref{fig:geophase3d} shows the graphical representation of
the geometric phase as a function of $\omega$ and $\epsilon$.
\begin{figure}[!htp]
	\centering
\begin{subfigure}[t]{.23\textwidth}
	\includegraphics[width=1.0\textwidth]{{{geophasexe}.pdf}}
	\caption{\small{Geometric phase $\gamma_{geo}(t_\omega)$ as a function of
	$\epsilon$ for fixed $\omega=2.0$ (full line), $5.0$ (dashed line) and
	$10.0$ (dotted line).}}
	\label{fig:geophasexe}
\end{subfigure}
\hspace{.01\textwidth}
\begin{subfigure}[t]{.23\textwidth}
	\includegraphics[width=1.0\textwidth]{{{geophasexw}.pdf}}
	\caption{\small{Geometric phase $\gamma_{geo}(t_\omega)$ as a function of
	$\omega$ for fixed $\epsilon=0.01$ (full line), $0.10$ (dashed line)
	and $0.40$ (dotted line).}}
	\label{fig:geophasexw}
\end{subfigure}
\caption{\small{Geometric phase plotted as a function of $\epsilon$ and
$\omega$.}}
\label{fig:geophase2d}
\end{figure}

We next consider two two-level quantum systems with individual
Hamiltonians given by \eqref{eq:h1}. When considering that the two systems
do not interact with each other, the phase factors obtained for the
composite system are simply the algebraic sum of the individual phase
factors. In order to explore how the phase factors of the composite
system change when interactions are taken into account, we considered an
interaction given by
\begin{equation}
	H'(t) = \kappa v(t) \ \sigma_3^{(a)}\otimes\sigma_3^{(b)},
\label{eq:hlinha}
\end{equation}
where $\kappa$ is a real constant and $v(t)$ is a real function of time.
The corresponding Hamiltonian in the rotated frame is given by
\begin{align}
	H'_2(t) &= R_y(\pi/2) \kappa v(t) \
	\sigma_3^{(a)}\otimes\sigma_3^{(b)} \adj{R_y}(\pi/2) \nonumber \\
	&= \kappa v(t) \ \sigma_1^{(a)}\otimes\sigma_1^{(b)}, \label{eq:hlinha2}
\end{align}
where $\epsilon_a$ and $\epsilon_b$ are the respective constants of the
individual systems and $f_a(t)$ and $f_b(t)$ are the external fields
applied to each subsystem. The Hamiltonian of the composite system is
\begin{equation}
	H_2(t) = \left( \begin{smallmatrix}
			f_a(t)+f_b(t) & \epsilon_b & \epsilon_a & \kappa v(t) \\
			\epsilon_b & f_a(t)-f_b(t) & \kappa v(t) & \epsilon_a \\
			\epsilon_a & \kappa v(t) & -f_a(t)+f_b(t) & \epsilon_b \\
	\kappa v(t) & \epsilon_a(t) & \epsilon_b & -f_a(t)-f_b(t)
	\end{smallmatrix} \right)
\label{eq:h2twointeracting}
\end{equation}

In order to obtain the phase factors for the composite system, we
consider the interaction picture. We will denote the state vector in this
picture by $\psi_I(t)$ and it relates to the state vector in the
Schrödinger picture by the unitary transformation
\begin{equation}
	\psi_I(t) = \adj{U}(t)\psi(t),
\label{eq:psii}
\end{equation}
where $U(t)$ is the time evolution operator. In the interaction picture, the time evolution
operator $U_I(t)$ is given by the Dyson series
\begin{equation}
	U_I(t) = \unit + \sum_{n=1}^\infty (-i)^n \int_0^t V_I(t_1) \dd t_1 \ldots
	\int_0^{t_{n-1}} V_I(t_n)\dd t_n.
\label{eq:ui}
\end{equation}
where $V_I(t)$ is the interaction Hamiltonian in the interaction picture
given by
\begin{align}
	V_I(t) &= \kappa v(t) \begin{pmatrix}
									V_{11}(t) & V_{12}(t) \\
									\conj{V}_{12}(t) & -V_{11}(t) 
								 \end{pmatrix}^{(a)} \otimes 
								 \begin{pmatrix}
									V_{11}(t) & V_{12}(t) \\
									\conj{V}_{12}(t) & -V_{11}(t) 
									\end{pmatrix}^{(b)}, \label{eq:vimatrix}
\end{align}
with
\begin{align}
	V_{11}(t) &= -\conj{U}_{11}(t)\conj{U}_{12}(t) - U_{11}(t)U_{12}(t),
	\label{eq:v11} \\
	V_{12}(t) &= \conj{U}_{11}(t)^2 - U_{12}(t)^2. \label{eq:v12}
\end{align}

The time evolution operator in the interaction picture given by the Dyson
expansion in \eqref{eq:ui}, considering the expression for the operator
$V_I(t)$ in \eqref{eq:vimatrix}, is
\begin{align*}
	U_I(t) &= \unit -i \kappa \int_0^t
	(\adj{U}_a(t')\sigma_1^{(a)}U_a(t'))  \\ 
	&\otimes ((\adj{U}_b(t')\sigma_1^{(b)}U_b(t'))\dd t' + \order (\kappa^2). 
\end{align*}
We shall consider the Dyson expansion up to first order.
The matrix form of
the time evolution operator in the interaction picture, in first order,
is given by
\begin{equation}
U_I(t) = \unit_4 -i\kappa V^{(1)}(t) ,
\label{eq:uimatrix}
\end{equation}
where  $\unit_4$ is the identity operator acting on a four-dimensional
Hilbert space and
\begin{equation}
	V^{(1)}(t)  = \int_0^t v(t') 
	\left(\begin{smallmatrix} 
			V_{11}^{(a)} V_{11}^{(b)} & V_{11}^{(a)} V_{12}^{(b)} &
			V_{12}^{(a)} V_{11}^{(b)} & V_{12}^{(a)} V_{12}^{(b)} \\
			V_{11}^{(a)} \conj{V}_{12}^{(b)} & -V_{11}^{(a)} V_{11}^{(b)} &
			V_{12}^{(a)} \conj{V}_{12}^{(b)} & -V_{12}^{(a)} V_{11}^{(b)} \\
			\conj{V}_{12}^{(a)} V_{11}^{(b)} & \conj{V}_{12}^{(a)} V_{12}^{(b)} &
			-V_{11}^{(a)} V_{11}^{(b)} & -V_{11}^{(a)} V_{12}^{(b)} \\
			\conj{V}_{12}^{(a)} \conj{V}_{12}^{(b)} & -\conj{V}_{12}^{(a)}
			V_{11}^{(b)} &
			-V_{11}^{(a)} \conj{V}_{12}^{(b)} & V_{11}^{(a)} V_{11}^{(b)}
	\end{smallmatrix} \right) \dd t'.	\label{eq:v1}
\end{equation}
We omitted the time-dependency of the expressions for
$V_{11}(t)$ and $V_{12}(t)$ given by equations \eqref{eq:v11} and
\eqref{eq:v12}, respectively. The operator $V^{(1)}(t)$ will be useful
for evaluating the expressions for the phase factors of the composite
system. Also, we must note that $V^{(1)}(t)$ is a self-adjoint operator,
since $v(t)$ is a real function of $t$ and the matrix operator in the
integrand on the right hand side of \eqref{eq:v1} is self-adjoint. 

The total phase factor for the composite system is
\begin{multline}
	\phi_{tot}(t) = \arg \pint{\psi_2(0)}{\psi_2(t)} \\
	= \arg	\Big\{\pint{\psi_2^{(a)}(0)}{U_a(t)\psi_2^{(a)}(0)}
	\pint{\psi_2^{(b)}(0)}{U_b(t)\psi_2^{(b)}(0)} \Big. \\
	 -i\kappa \pint{\psi_2(0)}
	{U(t)V^{(1)}(t) \psi_2(0)} \Big\}.
	\label{eq:totphaseinteract}
\end{multline}
Note that for $\kappa=0$ the expression above reduces itself to the
total phase of two non-interacting
systems.

Using \eqref{eq:dynphaset} for the dynamical phase and the expansion
in $\kappa$ for the time evolution operator in the interaction
picture, we have
\begin{align}
	\alpha_{dyn}(t) &= i\int_0^t \pint{\psi_2(t')}{\dot{\psi_2}(t')}\dd t'
	\nonumber \\
	&= i\int_0^t \pint{\psi_2(0)}{\adj{U}(t') \dot{U}(t') \psi_2(0)} \dd t'
	\nonumber \\
	&+ \kappa \int_0^t \pint{\psi_2(0)}{\adj{U}(t') U(t') V^{(1)}(t')
	\psi_2(0)} \dd t' \nonumber \\
	&+ \kappa \int_0^t \pint{\psi_2(0)} {\adj{U}(t') \dot{U}(t')
	\dot{V}^{(1)}(t') \psi_2(0)} \dd t' \nonumber \\
	&- \kappa \int_0^t \pint{\psi_2(0)} {\adj{V^{(1)}(t')}
	\adj{U}(t') \dot{U}(t')} \dd t' + \order(\kappa^2), \nonumber 
\end{align}
since $U(t)$ is unitary, the identity $\adj{U}(t)\dot{U}(t) =
-\dot{\adj{U}}(t)U(t)$ holds and the third term on the right hand side
of the expression above  can be rewritten as the complex conjugate of the
second term. Hence, the dynamical phase up to first order in $\kappa$ is
given by
\begin{align}
	\alpha_{dyn}(t) &= \alpha_{dyn}^{(0)}(t) \nonumber \\
	& + 2\kappa\Re 	\int_0^t \pint{\psi_2(0)}{\adj{U}(t') \dot{U}(t')
	V^{(1)}(t') 	\psi_2(0)} \dd t' \nonumber \\
	&\quad\quad  + \kappa \int_0^t \pint{\psi_2(0)} 
	{\dot{V}^{(1)}(t') \psi_2(0)} \dd t' + \order(\kappa^2),
\label{eq:dynphaseinteract}
\end{align}
where the $\alpha_{dyn}^{(0)}(t)$ is exactly the expression for the
dynamical phase for two non-interacting two-level systems. Also, the third term on the right hand side of
\eqref{eq:dynphaseinteract} is the integral over time of the expectation
value of the self-adjoint operator $V^{(1)}(t)$. Therefore, this term is
also real and so is the expression for the dynamical phase. The geometric
phase for the composite system is still given by the difference between
the total phase and the dynamical phase.

Now, let us consider the case in which the interaction is given by
\begin{equation}
	v(t) = \delta(t-t_0),
\label{eq:vdelta}
\end{equation}
where $t_0$ is any instant of time. The time evolution operator in the
interaction picture, according to \eqref{eq:uimatrix} and \eqref{eq:v1}, is
\begin{equation}
	U_I(t) = \unit_4 -i\kappa 
	\left(\begin{smallmatrix} 
			V_{11}^{(a)} V_{11}^{(b)} & V_{11}^{(a)} V_{12}^{(b)} &
			V_{12}^{(a)} V_{11}^{(b)} & V_{12}^{(a)} V_{12}^{(b)} \\
			V_{11}^{(a)} \conj{V}_{12}^{(b)} & -V_{11}^{(a)} V_{11}^{(b)} &
			V_{12}^{(a)} \conj{V}_{12}^{(b)} & -V_{12}^{(a)} V_{11}^{(b)} \\
			\conj{V}_{12}^{(a)} V_{11}^{(b)} & \conj{V}_{12}^{(a)} V_{12}^{(b)} &
			-V_{11}^{(a)} V_{11}^{(b)} & -V_{11}^{(a)} V_{12}^{(b)} \\
			\conj{V}_{12}^{(a)} \conj{V}_{12}^{(b)} & -\conj{V}_{12}^{(a)}
			V_{11}^{(b)} &
			-V_{11}^{(a)} \conj{V}_{12}^{(b)} & V_{11}^{(a)} V_{11}^{(b)}
	\end{smallmatrix} \right)_{t=t_0}, \label{eq:uidelta}
\end{equation}
where the time dependency of $V_{11}(t)$ and $V_{12}(t)$ are respectively
given by \eqref{eq:v11} and \eqref{eq:v12}. The time dependency in the second
term on the right hand side was omitted, but we assume that $0<t_0<t$
and so, both $V_{11}(t)$ and $V_{12}(t)$ are calculated for $t_0$, as is
indicated by the subscript on the matrix on the right hand side of
\eqref{eq:uidelta}.

\begin{figure}[!htp]
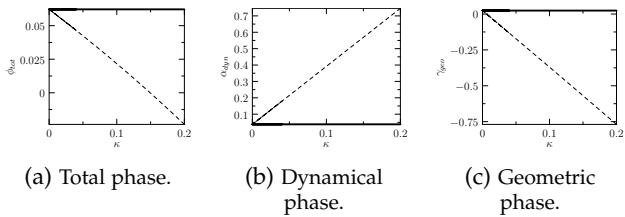

	\centering
	\begin{subfigure}[t]{.14\textwidth}
	\centering
		\includegraphics[width=1.0\textwidth]{{{totphase1000xk_w1_1.0_e1_0.01_w2_2.0_e2_0.01_t0_0.5}.pdf}}
		\caption{\footnotesize{Total phase.}}
		\label{fig:totphasexk}
	\end{subfigure}
	\hspace{.01\textwidth}
	\begin{subfigure}[t]{.14\textwidth}
	\centering
		\includegraphics[width=1.0\textwidth]{{{dynphase1000xk_w1_1.0_e1_0.01_w2_2.0_e2_0.01_t0_0.5}.pdf}}
		\caption{\footnotesize{Dynamical phase.}}
		\label{fig:dynphasexk}
	\end{subfigure}
	\hspace{.01\textwidth}
	\begin{subfigure}[t]{.14\textwidth}
	\centering
		\includegraphics[width=1.0\textwidth]{{{geophase1000xk_w1_1.0_e1_0.01_w2_2.0_e2_0.01_t0_0.5}.pdf}}
		\caption{\footnotesize{Geometric phase.}}
		\label{fig:geophasexk}
	\end{subfigure}
	\caption{\small{Plots of the phase factors for the initial state
			$\ket{00}$ as functions of the parameter $\kappa$. The thick line
			represents the value of the phase factors for a system with
			non-interacting subsystems. The dashed line represents the
	interaction given by \eqref{eq:vdelta}. We considered subsystems with
	$\omega_a=1.0$, $\omega_b=2.0$, $\epsilon_a=\epsilon_b=0.01$ and
	 $t_0=0.5$.}}
\label{fig:phasesxk}
\end{figure}

Up to first order in $\kappa$, the time evolution operator in
\eqref{eq:uidelta} is constant in time. Thus, the third term of the
expression for the dynamical phase in \eqref{eq:dynphaseinteract}, that
involves the time derivative of $V^{(1)}(t)$, is null. We implemented in
our code routines that calculate the phase factors for the interaction
given by \eqref{eq:vdelta}. To investigate the relation between the phase
factors and the constant $\kappa$, we considered a system composed of two
commensurable subsystems with fixed $\omega_a$, $\omega_b$, $\epsilon_a$
and $\epsilon_b$, a fixed $t_0$ that characterises the delta interaction
and we varied $\kappa$ from $0$ to $0.2$, with steps of $0.01$.
Considering this set of parameters, the code calculates the phase factors
for each of the computational basis states ($\ket{00}$, $\ket{01}$,
$\ket{10}$ and $\ket{11}$). Figure \ref{fig:phasesxk} shows the results
for the initial state $\ket{00}$ and $\omega_q=1.0$, $\omega_b=2.0$,
$\epsilon_a=\epsilon_b=0.01$ and $t_0=0.5$. The results are similar for
others sets of parameters. We note that since our
approximation of the Dyson expansion (equation \eqref{eq:ui}) is only up
to first order, the dependency of the phase factors on $\kappa$ is
linear. 
The parameter $\kappa$ is not, as one could imagine, a parameter of the
control space of the system. It simply modulates the interaction between
the subsystems and can be thought of as an structural constant.

\begin{figure}[!htp]
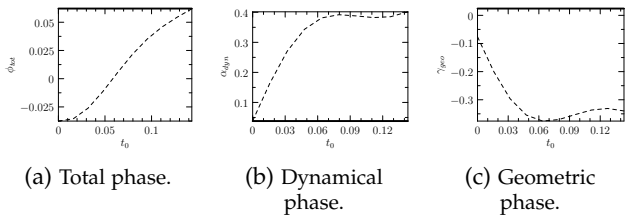

	\centering
	\begin{subfigure}[t]{.14\textwidth}
	\centering
		\includegraphics[width=1.0\textwidth]{{{totphase1000xt0_w1_1.0_e1_0.01_w2_2.0_e2_0.01_k_0.1}.pdf}}
		\caption{\footnotesize{Total phase.}}
		\label{eq:totphasext0}
	\end{subfigure}
	\hspace{.01\textwidth}
	\begin{subfigure}[t]{.14\textwidth}
	\centering
		\includegraphics[width=1.0\textwidth]{{{dynphase1000xt0_w1_1.0_e1_0.01_w2_2.0_e2_0.01_k_0.1}.pdf}}
		\caption{\footnotesize{Dynamical phase.}}
		\label{eq:dynphasext0}
	\end{subfigure}
	\hspace{.01\textwidth}
	\begin{subfigure}[t]{.14\textwidth}
	\centering
		\includegraphics[width=1.0\textwidth]{{{geophase1000xt0_w1_1.0_e1_0.01_w2_2.0_e2_0.01_k_0.1}.pdf}}
		\caption{\footnotesize{Geometric phase.}}
		\label{eq:geophasext0}
	\end{subfigure}
	\caption{\small{Plots of the phase factors for the initial state
			$\ket{00}$ as functions of the instant of time of the interaction
			$t_0$. The thick line represents the value of the phase factors
			for a system with non-interacting subsystems. The dashed line
			represents the interaction given by \eqref{eq:vdelta}. We considered
			subsystems with $\omega_a=1.0$, $\omega_b=2.0$,
			$\epsilon_a=\epsilon_b=0.01$ and $\kappa=0.1$. The time is measured
			in unites of $2\pi/\omega$.}}
\label{fig:phasesxt0}
\end{figure}

Figure \ref{fig:phasesxt0} shows the dependency of the phase factors on
the instant of time $t_0$ of the interaction for the initial state
$\ket{00}$. The presented relation between the phase factors and $t_0$ is
similar for the others states of the computational basis and for
different sets of parameters. We note that there is a value of $t_0$ that
maximises the absolute value of the geometric phase, but we cannot state
that this is a global maximum.

\section{Further results}

\begin{figure}
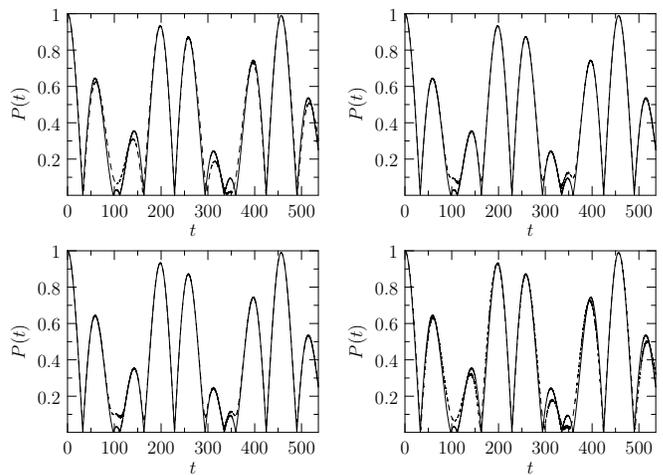

\centering
\begin{subfigure}{.23\textwidth}
	\includegraphics[width=1.0\textwidth]{{{p00to00_w1_1.0_e1_0.01_w2_2.0_e2_0.01}.pdf}}
\end{subfigure}
\hspace{.01\textwidth}
\begin{subfigure}{.23\textwidth}
	\includegraphics[width=1.0\textwidth]{{{p01to01_w1_1.0_e1_0.01_w2_2.0_e2_0.01}.pdf}}
\end{subfigure}
\begin{subfigure}{.23\textwidth}
	\includegraphics[width=1.0\textwidth]{{{p10to10_w1_1.0_e1_0.01_w2_2.0_e2_0.01}.pdf}}
\end{subfigure}
\hspace{.01\textwidth}
\begin{subfigure}{.23\textwidth}
	\includegraphics[width=1.0\textwidth]{{{p11to11_w1_1.0_e1_0.01_w2_2.0_e2_0.01}.pdf}}
\end{subfigure}
\caption{\small{Probability of the system remaining in its initial state.
Starting from the graph in the left column and first row, in clockwise
order the graphs correspond to the initial states $\ket{00}$, $\ket{01}$,
$\ket{10}$ and $\ket{11}$. The full line corresponds to non-interacting
subsystems and the dashed line corresponds to an interaction of the form
\eqref{eq:vdelta}. The time is measured in units of $t_\omega=2\pi/\omega$.
The relevant constants of the systems are $\omega_a=1.0$, $\omega_b=2.0$,
$\epsilon_a=\epsilon_b=0.01$, $\kappa=0.1$ and $t_0=.16\,t_\omega$.}}
\label{fig:transitions}
\end{figure}

Using the results obtained so far for two two-level quantum systems, we
may investigate once again the appropriate instant of time to calculate
the phase factors. Following the same prerogative, that the instant to
be considered corresponds to the time interval in which the system
undergoes a cyclic evolution, we consider the probability of transition
for the composite system:
\begin{equation*}
	P(t) = |\pint{\psi(0)}{U(t)\psi(0)}|^2.
\end{equation*}
Figure \ref{fig:transitions} shows $P(t)$ as a
function of time. We observe that the system returns to its initial state
after a time $T_\Omega\cong 456 t_\omega$, where $t_\omega=2\pi/\omega$.
$T_\Omega$ is also obtained through $T_\Omega=2\pi/\Omega$, where
$\Omega$ is the Rabi frequency and is calculated numerically. We
considered a system with $\omega_a=1.0$, $\omega_b=2.0$,
$\epsilon_a=\epsilon_b=0.01$. The constants that determine the
interaction are $\kappa=0.1$ and $t_0=0.5=0.16\,t_\omega$. For this
values, the correspondent Rabi frequency is $\Omega=0.0022$, resulting in
$T_\Omega\cong 456\,t_\omega$, as observed in Figure
\ref{fig:transitions}.

Once we determined the period that the system takes to return to its
initial state ($T_\Omega$), we can calculate the total phase factor of
the composite system. 
\begin{table}
	\centering
	\begin{tabular}{*7c}
		\hline
		{} & \multicolumn{2}{c}{$\omega_b = 1.0$} &
		\multicolumn{2}{c}{$\omega_b = 5.0$} & 
		\multicolumn{2}{c}{$\omega_b = 8.0$} \\ 
		\hline
		{} & $\phi_{tot}^{(0)}$ & $\phi_{tot}^{(\delta)}$ & 
		$\phi_{tot}^{(0)}$ & $\phi_{tot}^{(\delta)}$ & 
		$\phi_{tot}^{(0)}$ & $\phi_{tot}^{(\delta)}$  \\
		\hline
		$\ket{00}$ & 0.027 & -0.116 & -1.816&  -1.878 & -2.491 & -2.552 \\
		$\ket{01}$ & 0.000 & 0.151 & 1.843 & 1.898 & 2.518 & 2.570 \\
		$\ket{01}$ & 0.000 & 0.151 & -1.843 & 1.790 & -2.518 & -2.466 \\
		$\ket{00}$ & -0.027 & -0.116 & 1.816 & 1.752 & 2.491 & 2.429 \\
		\hline
	\end{tabular}
	\caption{\small{Values of the total phase $\phi_{tot}$ for each state
			of the computational basis, considering different
			values of $\omega_b$ and fixed $\omega_a=1.0$. The subscript
			$\phi_{tot}^{(0)}$ and $\phi_{tot}^{(\delta)}$ indicate systems
			with no interaction and interaction given by a delta function,
			respectively.}}
	\label{tab:totvalues}
\end{table}

Table \ref{tab:totvalues} shows values of the total phase for a set of
$\omega_a$ and $\omega_b$ values. We note that when $\omega_a=\omega_b$,
we can write the following transformation:
\begin{equation}
	B(\phi) = \left( \begin{array}{cccc} e^{i\phi} & 0 &
	0 & 0 \\ 0 & 1 & 0 & 0 \\ 0 & 0 & 1
		 & 0 \\ 0 & 0 & 0 & e^{-i\phi} \end{array} \right),
	\label{eq:bbarata}
\end{equation}
where $\phi$ is the total phase associated with the basis state
$\ket{00}$. This transformation implements a conditional evolution of the
basis states, we can say that \eqref{eq:bbarata} is a conditional phase
gate in the sense that the state of one system influences the state of the
other, although it does present the usual symmetric form of controlled
phase shift gates. This gate is not purely geometrical, since the total
phase factor involves both the dynamical and geometric phases. When
$\omega_a\neq\omega_b$, the transformation on the basis state can no
longer be represented by \eqref{eq:bbarata}, as can be seen in Table
\ref{tab:totvalues}.

\section{Conclusions}

The main contribution of this work is the implementation of the method
developed in \cite{ref:barata} and \cite{ref:cortez} to obtain phase
factors for a two-level quantum 
system and two two-level quantum systems interacting and non-interacting.
Since this method presents a solution stable for long-time periods,
the resulting phase factors also present this property. 

The implementation of a quantum gate, when RWA is considered
\cite{ref:ekert, ref:sjoqvist} is valid for an
adiabatic evolution and, in the context of two two-level systems
interacting, only one is subjected to an external time-dependent field.
In our case, both systems are subjected to an external periodic field and
neither the adiabatic approximation nor the rotating wave approximation
are necessary. Using the results for phase factors we were able to
implement a controlled phase shift gate. The resulting gate is not purely
geometrical and removing the dynamical contribution to the overall phase is not a
straightforward task. One possibility is finding a Hamiltonian that
cancels the dynamical phase of the system along a cyclic trajectory.
Nevertheless, our work can be extended in many ways. For example, the
time evolution operator obtained for a two-level quantum system could be
used in the calculation of geometric phases in open quantum systems
under the Quantum Jump Approach \cite{ref:carollo}. Or, for non-unitary
evolutions in the context of interferometry, it is even
possible to combine the method developed in \cite{ref:peixoto} with our
work to obtain a time evolution operator for a system subjected to a
time-dependent perturbation and derivate the corresponding phase factors.

 \bibliographystyle{plain}
 \bibliography{bib}

\end{document}